# The COMPASS Event Store in 2002

V. Duic
*INFN, Trieste, Italy*

M. Lamanna
*CERN, Switzerland and INFN, Trieste, Italy*

COMPASS, the fixed-target experiment at CERN studying the structure of the nucleon and spectroscopy, collected over 260 TB during summer 2002 run. All these data, together with reconstructed events information, were put from the beginning in a database infrastructure based on Objectivity/DB and on the hierarchical storage manager CASTOR. The experience in the usage of the database is reviewed and the evolution of the system outlined

## 1. INTRODUCTION

COMPASS (COmmon Muon Proton Apparatus for Structure and Spectroscopy) [1] is a fixed-target experiment with an extensive physics programme at the CERN SPS using different configurations, notably using both muon and hadron beams in the 100-300 GeV range at very high intensities. The COMPASS experiment started taking data in Summer 2001; in Summer 2002 the first full physics run was performed, collecting data for the measurements of ΔG/G of the Deuteron and to investigate transversity effects.

In this paper we will mainly focus on the experience of the first year of physics data acquisition, comparing the design figures with the actual achievements and elaborating a roadmap for the future. The value of COMPASS experience in computing is not limited to the experiment itself since it is a field experience in key LHC technologies.

## 2. COMPASS OFF-LINE SYSTEM

The off-line system was built to meet severe design constraints, namely the continuous high data acquisition rate (about 40 MB/s) and the very large data sample ($10^9$ events, 30 kB each, 300 TB/year) to be reconstructed virtually on-line.

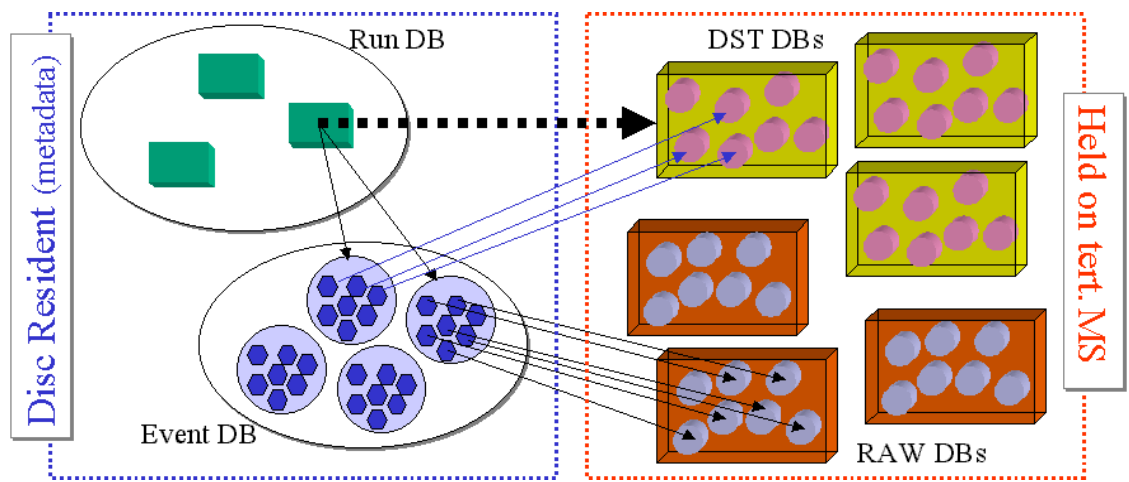

Figure 1. Database Access and Storage Schema.

The large volume of data to be processed (during a few months each year) and the need for a flexible software environment to cope with the different configurations and measurements of the experiment pushed the COMPASS Collaboration to design the off-line analysis software from scratch and to build a dedicated facility for the off-line computing, namely the COMPASS Computing Farm (CCF) [2].

COMPASS decided to build a Central Data Recording (CDR) System[1] to record all the data: the on-line system does not write the events on tape at the experiment site, but sends them over a few kilometres of dedicated optical fibre network to the computer centre, where the CCF, the tape servers, and the corresponding high-speed tape drives are located.

---
[1] It is based on the original CERN CDR, but implemented and managed by COMPASS.

In 2002, all the data were entered in a database infrastructure based on Objectivity/DB: the original event files are the input of a procedure that populates the database hierarchically, to ease the navigation during reconstruction and analysis. The main feature of the database implementation is that to each event corresponds a small object called header keeping all the basic information on that event, including the RAW data database pointer to the RAW event and multiple DST versions (reconstructed objects).

The main access to the data (Figure 1) goes via these header objects, keeping the possibility to connect at run time all the components of an event (for example, the association between the DST and the RAW event is kept on a event by event basis). The headers are always on-line, while the files containing the RAW and DST events are controlled by the CERN Hierarchical Storage Manager (HSM) CASTOR [3] via a modified AMS, the Objectivity/DB object server (CERN IT/DB). Access to full events is therefore made transparent: the possible cases of a programme accessing local data, data on a file on a remote server, or recalling data from tape are not exposed in the user software.

The estimated computing power to reconstruct all the events at the speed of the data acquisition is 20,000 SI2K; the actual figure has grown to about 100,000 SI2K in the last versions of the reconstruction programme, which can be provided by some 100 Linux Dual CPU PCs. The choice for the network technology is Gigabit and Fast Ethernet. A disk pool of a few TB has been set up, initially made up of SCSI disks, but more recently using the less expensive EIDE disks.

### 3. 2002 COMPASS DATA TAKING

In 2002, COMPASS has been taking data for 12 weeks: after a first phase of detector and data acquisition commissioning, the rate from the experiment was consistently higher than the design value. Peaks over 4 TB of database files to tape have been observed, corresponding to databases actually moved to tape (Figure 2).

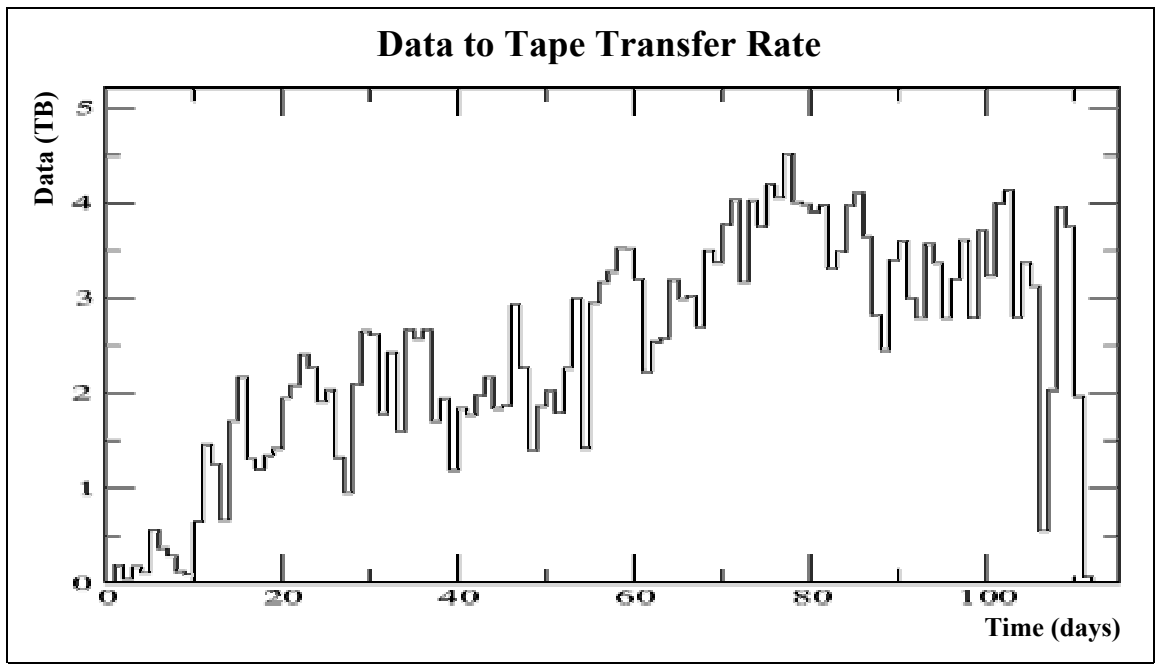

Figure 2. Total size of database files (in TB) copied to CASTOR as a function of time (days).

In a nutshell, the on-line system groups the events in files (1 GB file maximum). These files flow in parallel multiple streams to the off-line system via RFIO protocol [3] to the CCF. The events from the transferred files are input in a federated database and then registered in the CASTOR name space: from this moment onward, CASTOR controls completely the events data handling (copy to tape, managing of the disk space, recall of databases present only on tape). In every stage of the storage pipeline, the system maintains the original files in some buffer areas on both the on-line and the off-line farms, and deletes them as late as possible after a successful copy to tape has been produced.

### 4. EXPERIENCE WITH OBJECTIVITY/DB

The strong points COMPASS has observed in operating large Objectivity/DB federations were the following:
- The separation between the logical and the physical layer assured independence from the HSM details;

- Thanks to the central role and the straightforward manageability of the catalogue as a repository for databases physical location handling, it has been possible to integrate the product with the HSM in a clean way; in particular, it was possible to balance in a transparent way the load on the different machines serving as the CASTOR disk cache;
- The client-driven nature of Objectivity/DB has shown to be pretty suited to the access pattern of COMPASS data handling, which has resulted in the possibility of achieving a high degree of concurrence in production (up to concurrent 400 clients); the AMS is designed to deal with many concurrent clients via a set of lightweight stateless processes.
- A simple but effective file protection (against accidental errors) could be put in place in the CERN version of the AMS (purely based on user UID). This was sorely needed since Objectivity/DB tools do not implement any user access policies management.

On the other hand, the live experience with that ODBMS has lead COMPASS to deal with the following drawbacks:

- Many transactions were not automatically rolled back in case of common client failures: the recovery from this situation was a major burden for the database administrators. To ease the clean up, the generic users were accessing the federation in read-only mode. This feature is implemented via the reconstruction programme, whilst a centralised ACL system would have been very useful.
- The clean up of aborted transactions came out to be really complex; in some cases, a restart of the Lock Server was the only viable solution, which on the other hand lead all user applications to abort. The removal of locks becomes a problem above 100 clients accessing the database.
- Objectivity/DB Lock Server and AMS can be effectively operated as "1 box per process and per database federation" but the available COMPASS resources compelled the collaboration to make some compromise to this requirement.
- The database creation is a particularly delicate operation that exhibits an imperfect "atomic" nature, and might lead to update locks that lock the whole federation. The main workaround was to create the databases via a shell script using the Objectivity/DB line command *oonewdb* instead of the corresponding Objectivity/DB API; this almost solved the problem, nevertheless during data taking in a few cases similar locks were observed. These locks are extremely dangerous because they prevent any data from being written (COMPASS can continue to take data thanks to the CDR multiple buffers; once the lock is removed, the CDR can catch up by sending data to tape at higher speed). The resolution of these problems does require expert intervention, which is unlikely to be automatizable because it relies on unpublished options of Objectivity/DB tools. The regime where these problems are likely to appear is well above 40 MB/s, and a few database created per minute (in this situation the central infrastructure holding the database catalogue is 100% busy at dealing with these operations).
- Objectivity/DB introduces an overhead in the database files of ~30% of the RAW data size; thanks to compression, once on tape, it translates into a factual ~6% overhead (compared to the corresponding compressed size of the RAW data); this effect was known, but it has to be kept in mind because of financial consideration on the recording media costs.

## 5. 2002 DATA PRODUCTION

As the technical run has taken place in 2001 (collecting few tens of TB), COMPASS has managed to submit to production those data in Spring 2002. The 2002 data have been first processed during the run (small samples) whilst a larger production has been carried on in the subsequent months; the data production was stopped when the migration of the entire data store to Oracle 9i was started (see below).

As far as 2001 data is concerned, the amount of produced DSTs is relatively modest: only 103 GB have been reconstructed, but these data allowed studying the apparatus and looking for basic physics signal. The production involving 2002 data is characterised by much bigger figures:

- 5.7 TB of DSTs have been produced out of 80 TB of RAWs under Objectivity/DB.
- 4 TB of DSTs have further come out, up to the date of this review, from the reconstruction of 80 TB of the RAW data under Oracle (production still going on).

As Objectivity/DB will be finally dismissed by the end of May 2003, the analysis of the DSTs still residing on that ODBMS will have to end before that deadline, though it still will be possible to access those data from within any dedicated environment of the supported OS platform at any time (Linux Red Hat 6).

In fact, the analysis task in COMPASS is in charge to the participating institutes scattered across the world, which have already locally transferred part of the available DSTs of their interest and which have not been provided yet with the new DMBS. Up to now, thanks to the manageability of Objectivity/DB, some institutes have been able to cope with it by themselves. For example, INFN – Trieste has set-up a shallow copy of the 2002 federation, provided with some 1.4 TB of the DST data, on which they are being performing their analysis using the satellite computing Farm – ACID [4].

Other participating institutes perform their analysis directly on mini DSTs, which are selectively filtered out from the DSTs (their turn out to have a size of about 1 GB per run – approximately 1% of the original RAW data). They are stored as ROOT files (no database infrastructure is used to access them), and the analysis is performed by means of PHAST [5], the COMPASS framework for the data analysis at mini DST level, which:

- Provides access to reconstructed events;
- Is the environment for final analyses;
- Is a tool for further filtering out sub-samples of events.

## 6. MIGRATION TO ORACLE

Since CERN has decided to terminate the support contract for Objectivity/DB by Objectivity Inc, COMPASS has been forced to adopt a new solution for its data storage: Oracle 9i has been chosen to be the new DBMS.

The technology choice and the implementation (devised at CERN by the IT/DB group and validated by COMPASS) take into account the two main lessons learnt in building and operating the previous system. First of all, Oracle 9i as an infrastructure shows very good feature of error resiliency, making the interaction with it relatively simple from the user point of view; the CERN IT/DB experience in running Oracle-based services plays also an important role. The second lesson is that once the meta-data are in the database, the bulk of the data could be addresses simply via filename and file offset (this was also the back-up solution for the Objectivity/DB implementation before demonstrating its scalability properties).

The new meta-data corresponding to the 2002 data yields about 500 GB.

Unlike for Objectivity/DB, COMPASS is not primarily contributing to the management of the database infrastructure, which is provided exclusively by CERN IT/DB division.

No direct DST database migration has been performed, since COMPASS has profited from the DBMS migration to improve the format of DSTs, and to improve on calibrations and algorithms.

### 6.1. Oracle tests

The migration of the COMPASS data has been performed by the CERN/DB group, and described at this conference [6]; it has been validated by a series of checks aimed at verifying that the new system was correctly set-up and working, and tests carried on in order to have a preliminary picture of the new system, as soon as a minimal number of RAW data were migrated.

Consistency checks have been performed by COMPASS on basically 100% of the data for the first samples migrated (comparing each event from the Oracle 9i and the Objectivity/DB stores). Globally, about 5% of the data have been selected randomly across the full sample to verify data integrity.

The functionality of the new store has been verified by joint activities of COMPASS and the CERN IT/DB group.

The stress tests that have been undertaken consisted in accessing the RAW data for reading by a variable number of concurrent clients. The tests were repeated for different numbers of concurrent clients while the concerned storage system components (Oracle 9i, and RFIO daemons), CPU and network loads were monitored, to gain information on how the access was stressing the system.

This kind of access to the data reproduces the typical access pattern in COMPASS data handling, though a simultaneous start of many clients is explicitly introduced to stress the system as much as possible.

As far as the users point of view is concerned, the performance in scanning the data contents, measured by time spent to scan a certain amount of data, as a function of the concurrent accessing clients, we observed a very nice scalability of the implemented system, up to about 100 concurrent clients (Figure 3).

Besides the scalability properties, the new system has also shown to be able to provide the clients with the requested data at a speed that was as close as the network bandwidth could afford so that the new system did not show any limitation arising from the server itself, unlike we had observed for Objectivity/DB.

Also the absolute performances were better for Oracle 9i than for Objectivity/DB.

It is worth saying that, independently of the better performances of the new system, the one based on Objectivity/DB was anyway well suited to the access requirements both in terms of aggregate rate and concurrency. In fact, the design data flow rate that rules the CCF activity during CDR is set to 35 MB/s, well below the rates obtainable by 10 concurrent clients (which is below the saturation affecting Objectivity/DB).

The further information drawn from monitoring[2] the system, demonstrated that, thanks to the adoption of the hybrid system mentioned above, the data access load is shared between the DBMS and the RFIO daemons because, as soon the Oracle server locates the requested data-file (and relevant data-block entry point), the data intensive transfer is performed by a RFIO daemon on the data server hosting that file.

The very good performances observed in scanning large sets of data are inherited from the RFIO system which is optimised for this use case; to every active client corresponds an RFIO daemon serving data over the network.

The only limitations we have observed so far are the following.

---

[2] By the monitoring tools and methodologies designed for and applied by COMPASS to its CDR system [2].

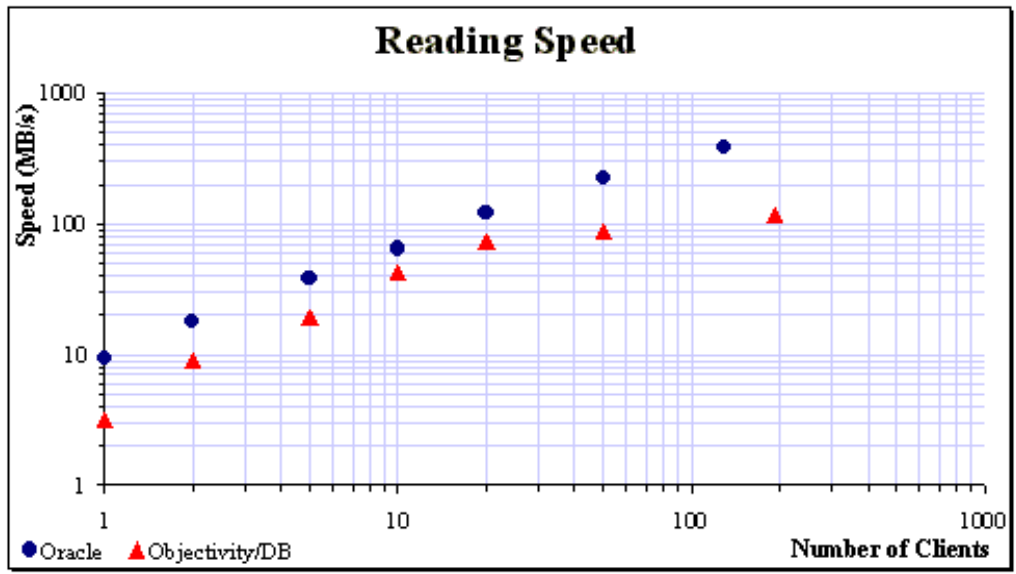

Figure 3. Comparison between Objectivity/DB and Oracle 9i based storage systems when accessing COMPASS RAW data for reading. The new system (closed points) shows a better performance than the one based on Objectivity/DB (closed triangles).

- Queries selecting a large number of events tend to overload the server (in DST reading, for example a common query selects a few millions Oracle records); splitting such a kind of selections in multiple queries to optimise the performances will solve this problem.
- Furthermore, the better performance of the RFIO daemon (compared to the AMS thread/sub-process of Objectivity/DB) can be cancelled if a single data server is trying to serve data to too many clients (due to the relative heavier structure of RFIO in terms of required system resources). This issue was avoided by a properly designed data-server load in running conditions during the production of early 2003, where a set of less than 10 data servers could sustain over 400 clients reading at an aggregate speed of more than 20 MB/s, and writing approximately 2 MB/s of data to tape.
- The load balance across servers has to be done (in the present infrastructure) via a load balancing system residing in a library linked to the user code; in the Objectivity/DB store, this parameter was instead explicitly expressed as a property of the databases, therefore independent from any user application.

## 7. CONCLUSION AND OUTLOOK

As shown in Figure 4, COMPASS has entered in the second big production phase. The first phase (late 2002) corresponds to the access of 2002 data via Objectivity/DB. As calibrations and alignment have been improved, the new production is a full reprocessing of 2002 data, using the Oracle 9i infrastructure. The latter production should now get more momentum and should be finished by the end of summer. The CDR of 2003 data (using Oracle 9i) has just started, and it will be running in parallel with the current DST production and with occasional further access to the new data (the planned on-line reconstruction of the data is still not viable due to the long time of preparation for the calibration sets: if these constants cannot be calculated within few hours, the on-line processing cannot benefit from the fact that the CDR data are cached for about 24 hours on the CCF disks, requiring a tape recall also for the "new" data).

The 2003 CDR is expected to populate a data store with over 200 TB of new data in the period May - September 2003.

COMPASS will continue to play an important role as user of the computing infrastructure at CERN in many different areas:

- Computing fabric – COMPASS computing power needs are (in production mode) the highest at CERN for analysing real data, rivalling with the large computing challenges of the LHC experiments;
- Data management – the COMPASS data store integrated size is already the largest at CERN [3];
- Central Data Recording – COMPASS has the highest rate to mass storage with physics data among the current CERN experiments, close to the expected rate for the multipurpose LHC experiments.

Moreover, the extensive use of different approaches to data management has yielded very interesting sets of experience as a result of:

- The deployment of multiple complex technologies for the data management used in production;

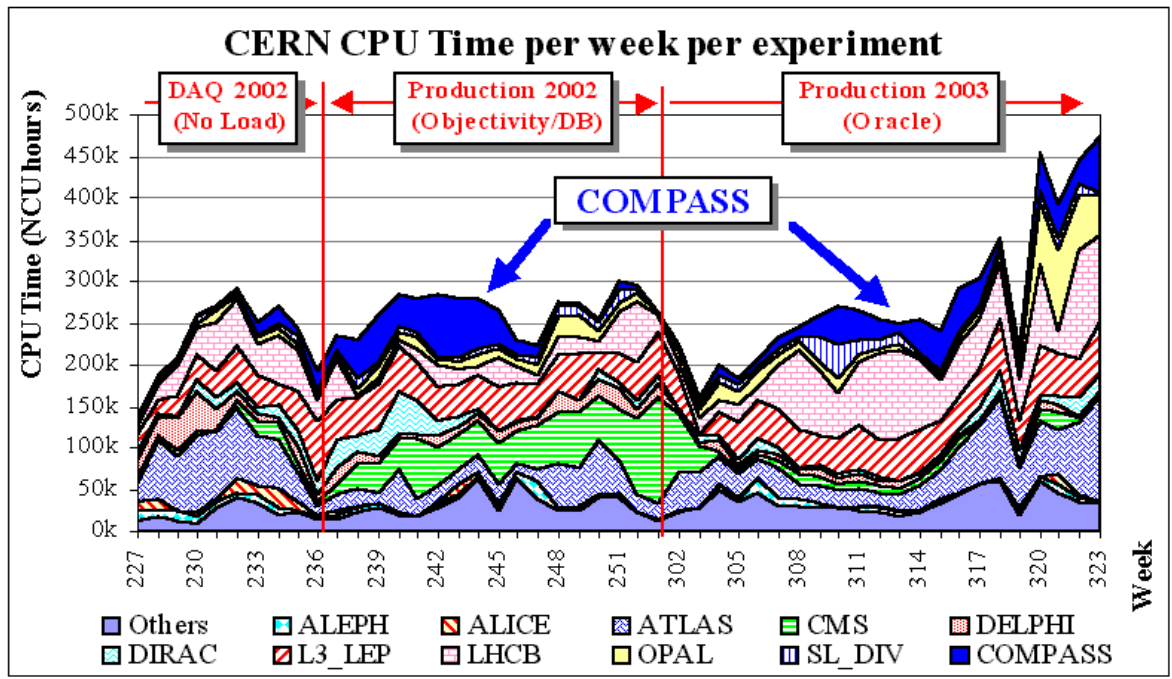

Figure 4. Batch CPU load at CERN. The COMPASS CPU usage is outlined.

- The operation of very large data stores of real data;
- The successful change of a database technology within an experiment and the corresponding data migration, with no disruption in the continuity of data access.

All this experience has contributed in building a solid system for COMPASS, and, as a by-product, a fertile exchange of technological and operational experience at CERN.

## Acknowledgements

The authors would like to acknowledge the encouragement and support of the COMPASS collaboration and of the CERN IT division. The authors would like to explicitly thank the CASTOR team (in particular J. D. Durand), whose constant and enthusiastic support has made it possible to effectively operate such a complex system.